\documentclass[twocolumn,showpacs,preprintnumbers,amsmath,amssymb]{revtex4}
\usepackage{bbm,amsfonts,amssymb,epsfig,verbatim,mathbbol}

\begin{document}
\title{On the zero of the fermion zero mode}
\author{Falk Bruckmann}
\affiliation{Instituut-Lorentz for Theoretical Physics, University of Leiden (NL)}
\begin{abstract}
We argue that the fermionic zero mode in non-trivial gauge field backgrounds must
have a zero.
We demonstrate this explicitly for calorons where its location is related to a
constituent monopole.
Furthermore a topological reasoning for the existence of the zero is given
which therefore will be present for any non-trivial configuration.
We propose the use of this property in particular for lattice simulations in
order to uncover the topological content of a configuration. 

\end{abstract}
\maketitle
 
\def\P{\mathcal{P}_\infty}
\def\Tr{{\rm{Tr}}}
\def\D{{\rm{D}}}
\def\p{\Psi_z}
\def\pp{\hat{\Psi}_z}
\def\po{\Psi_0}
\def\ppo{\hat{\Psi}_0}
\def\F{\mathcal{F}}
\def\f{\hat{f}_x}
\def\v{\!\!\!\!\!\!\!\!\!\!\!\!\!\!\!}

\section{Introduction}

Fermionic zero modes reflect properties of the background gauge field in a very
interesting manner.
For a gauge field with topological charge $k$ the index theorem guarantees that
the difference of the number of left-handed and
right-handed zero modes equals $k$, which in the generic case means 
 $|k|$ zero modes of definite chirality.

The maxima of the zero mode density generically agree with the maxima
of the action or energy density. This localization property is well-known for the
charge 1 instanton resp. the charge 1 magnetic monopole and holds for higher
charge as long as the individual building blocks are well separated.

The zero mode also acts as a filter, i.e. its profile is fairly smooth
even when the background gauge field is rough.
The reason is that the vanishing energy eigenvalue does
not allow for large (gauge covariant) momenta.

All these statements also apply to calorons, instantons at finite temperature,
i.e.\ over $S^1\times \mathbb{R}^3$.
At non-trivial holonomy calorons dissociate into constituent monopoles
which made these objects attractive over the last years.
The dilemma of the zero mode
(resp. the $|k|$ zero modes) where to localize to
is solved in the following way:
the zero mode localizes to one (type of) monopole, which one
depends on the phase in the fermion boundary condition in the time-like direction.
These boundary conditions are meant as a diagnostic tool,
physical fermions are antiperiodic.
This characteristic hopping has also been demonstrated for
equilibrated configurations \cite{gattringer:02b},
even at zero temperature, i.e.\ on the four-torus \cite{gattringer:04a}.

For obvious reasons attention so far was paid to the maximum of these zero
modes.
In this letter we analyze another interesting feature of the zero mode namely
that it has a zero.
\begin{comment}
In this letter we show that the zero mode has another interesting feature,
namely a {\em zero} at one point in space-time.
We demonstrate this novel property explicitly for calorons 
and show that it is related to a constituent monopole.
We explain the topological origin of the zero which hence persists
when leaving classical solutions
(but staying in a non-trivial sector).
Then we generalize our results to higher topological charge and compact space-times.
Finally, we comment on applications of our findings to gauge fixing and lattice
simulations.
\end{comment}
In order to describe the main effects, we will focus on the gauge group $SU(2)$.

\subsection{Calorons in brief}

Compared to the Harrington-Shepard solution \cite{harrington:78} the more recently
found calorons \cite{kraan:98a,lee:98b} have a new aspect:
the non-trivial holonomy.
It means that the Polyakov loop at spatial infinity $\P$
(being direction-independent for finite action)
does not necessarily equal $\pm \Eins_2$.
Quite the contrary, calorons with holonomy close to maximally non-trivial,
$\Tr\,\P=0$, shall be of relevance for the confined phase
where this property holds on average.
In the following we will give the main properties of charge one calorons
and their zero modes, for a review see \cite{bruckmann:03c}.

The most fascinating feature of non-trivial holonomy calorons is the fact
that they dissociate into a monopole-antimonopole pair 
(a charge $k$ caloron can be described by $|k|$ monopoles and $|k|$
antimonopoles).
They show up in a gauge independent way as (anti-)selfdual lumps of action density.
The holonomy
acts like a Higgs field and hence determines the masses of the monopoles. For
maximally non-trivial holonomy all monopoles are of the same mass,
while for the HS solution all but one types of monopole are massless and thus
invisible. 

The other mechanism at work is dimensional reduction,
in the sense that the caloron size $\rho$ competes with the extension of the compact
direction $\beta=1/k_B T$ (which will be set to 1 in the following).
For large $\rho$ the caloron consists of two almost static monopole lumps with distance
$\pi\rho^2$, while the action density of small calorons looks like that of small
instantons.

The winding number of the Polyakov loop and its non-trivial value at infinity imply
that there are spatial locations where the Polyakov loop passes
through $\Eins_2$ and $-\Eins_2$, respectively.
They provide an alternative definition of the monopole locations since
in the core of a monopole the broken symmetry is restored. 
Interestingly, the Polyakov loop shows this dipole even for small instantons
(inside the single lump of action density) which distinguishes small calorons
from instantons.

\section{The zero of the caloron zero mode}

Chiral zero modes in the background of a (caloron) gauge field fulfill
\begin{eqnarray}
\bar{\sigma}_\mu (\partial_\mu+A_\mu)\p(x)=0\,,\qquad
\bar{\sigma}_\mu=(\Eins_2,-i\tau_i)\,,
\label{eqn_eqn}
\end{eqnarray}
with $\tau_i$ the Pauli matrices and boundary conditions
\begin{eqnarray}
\p(x_0+1,\vec{x})=e^{-2\pi i z}\p(x_0,\vec{x})\,.
\end{eqnarray}
The (antihermitean) gauge fields $A_\mu$ are periodic in $x_0$.
Alternatively one can make the zero mode periodic upon gauge transforming
with $e^{2\pi i z x_0}$, whereby a mass term $2\pi i z$ is added to $A_0$.

The dependence of $\p$ on the boundary conditions can be best understood
in the regime of well separated constituents which then become BPS monopoles. 
According to the Callias index theorem \cite{callias:77}
the latter possess fermion zero modes
for given $z$-ranges, since the effective mass governs the asymptotic behavior
of the modes which is an exponential decay.
The monopoles enter the caloron in such a way that
the zero mode of the caloron hops from one 
constituent monopole to the other upon changing $z$. 

More precisely, choosing the holonomy to be diagonal,
\begin{equation}
\mathcal{P}_\infty=\exp(2\pi i \omega \tau_3)\,,\qquad
\omega\in [0,1/2]\,, \label{eqn_holonomy}
\end{equation}  
the monopole masses are $8\pi^2\nu_m,
\nu_1=2\omega,\nu_2=2\bar{\omega},\bar{\omega}=1/2-\omega$, respectively.
The zero mode $\p$ with $z\in(-\omega,\omega)$ localizes to the first monopole,
while for $z\in(\omega,1-\omega)$ it localizes to the second one.
At the boundaries $z=\pm \omega$ the zero mode sees both constituents and is 
delocalized.

It has been shown that in the well separated regime the fermion zero mode 
around its maximum becomes the spherically symmetric BPS zero mode
\cite{garciaperez:99c}. 

In the following we will show that the `invisible' monopole still distorts 
the zero mode such that the latter develops a zero
in the vicinity of that monopole.

\enlargethispage{\baselineskip}

For simplicity we choose the monopoles to lie on the $x_3$-axis with their centers at
$y_1=-\nu_2\pi\rho^2$ and $y_2=\nu_1\pi\rho^2$,  
such that the center of mass is always at the space-time origin.
For the distance to these locations we introduce
\begin{eqnarray}
&&r_m=|\vec{x}-\vec{y}_m|\,, \quad
s_m=\sinh(2\pi\nu_m r_m)\,, \nonumber\\
&&c_m=\cosh(2\pi\nu_m r_m)\,, \quad
m=1,2,\:\:\mbox{no sum.}
\end{eqnarray}
The analytic formula for the zero mode is based on the ADHM-Nahm formalism. With
color index $a$ and spin index $\alpha$ it reads
\begin{eqnarray}
\p(x)\!\!\!\!\!\!&&_{a\alpha}=\frac{\rho}{2\pi}\sqrt{\phi}\,
\exp(-2\pi i\omega x_0\tau_3)_{ab}\nonumber\\
&&\partial_\mu\left(\f(-\omega,z)P_1\bar{\sigma}_\mu+
\f(\omega,z)P_2\bar{\sigma}_\mu\right)_{b\beta}
\epsilon_{\beta\alpha}\,.
\end{eqnarray}
For our choice (\ref{eqn_holonomy}) the projectors are
$P_{1,2}=(\Eins_2\pm\tau_3)/2$ and
$\phi=(1-\rho^2\hat{f}_x(-\omega,-\omega))^{-1}$ is a positive function.

\begin{figure}[b]
\includegraphics[height=2.5cm]
{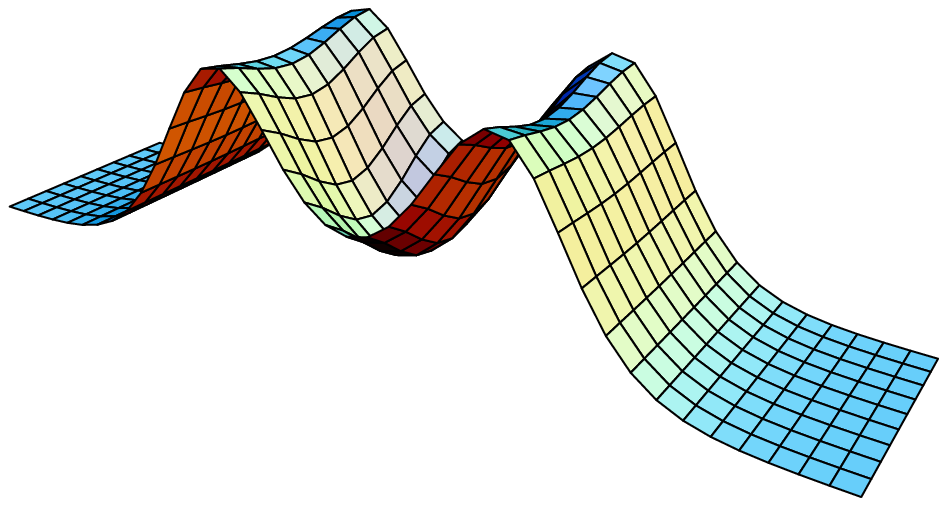}
\hspace*{-1.1cm}
\includegraphics[height=2.5cm]
{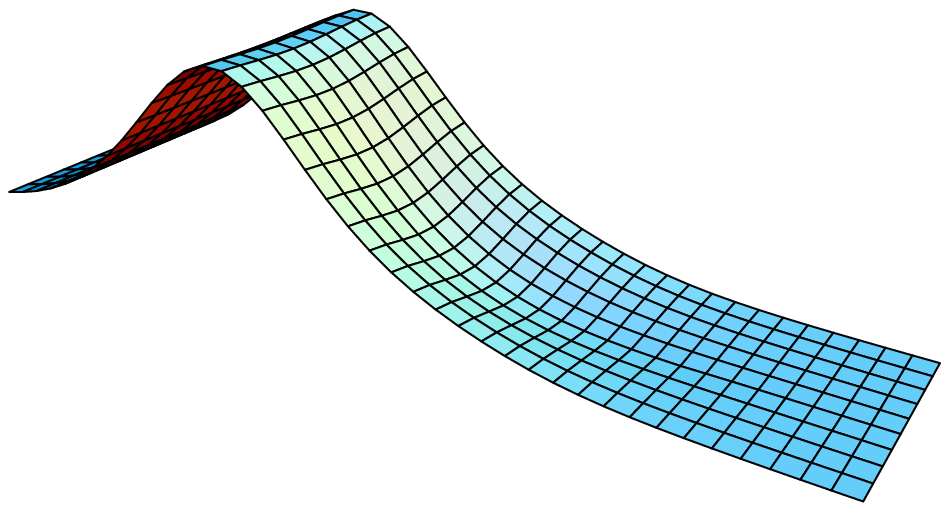}
\caption{Action density (left) and periodic zero mode (right) for a caloron of
size $\pi\rho^2=1.40$ and holonomy $\omega=1/4$ (equal mass case) in the $x_0x_3$
plane.
The antiperiodic zero mode can be obtained from the periodic one by reflecting
$x_3$ around $0$. 
}\label{fig_first}
\end{figure}

Without loss of generality we will concentrate on close to periodic zero modes,
$z\in(-\omega,\omega)$, which localize to the monopole at $y_1<0$
(the other zero modes can be obtained via an antiperiodic gauge transformation
which interchanges $\nu_1\leftrightarrow\nu_2$ and thus
$y_1\leftrightarrow -y_2$
\cite{garciaperez:99c}).
An impurity scattering formalism for $\hat f_x$ yields then \cite{bruckmann:02b}
\begin{eqnarray}
\f(-\omega,z)&=&-4\pi^2\hat{g}^\dagger(-\omega-z)
\left\{\left[(\Eins_2-\F_{-\omega})^{-1}-\Eins_2\right]\right.\nonumber\\
&&\left.W^{-1}(z+\omega)\right\}_{12}\,,
\end{eqnarray}
where the index 12 refers to the upper right entry and
\begin{eqnarray}
\v&&\hat{g}^\dagger(z)=\exp(2\pi i x_0 z)\,,\qquad
\F_{-\omega}=\hat{g}^\dagger(1)TH_2TH_1\,,\nonumber\\
\v&&H_m=\left(\begin{array}{cc}
c_m&s_m/2\pi r_m\\
2\pi r_m s_m&c_m
\end{array}\right),\qquad
T=\left(\begin{array}{cc}
1&0\\
2\pi^2\rho^2&1
\end{array}\right),\nonumber\\
\v&&W^{-1}(z)=\!\left(\!\begin{array}{cc}
\cosh(2\pi r_1 z)&\!\!-\sinh(2\pi r_1 z)/2\pi r_1\\
-2\pi r_1\sinh(2\pi r_1 z)&\!\!\cosh(2\pi r_1 z)
\end{array}\!\!\right).
\end{eqnarray}
The other Green's function $\f(\omega,z)$ can be obtained by 
\begin{eqnarray}
\f(z,z')=\f^*(-z,-z')\,.\label{eqn_fprop}
\end{eqnarray}

\begin{figure}[b]
\includegraphics[height=3.5cm]{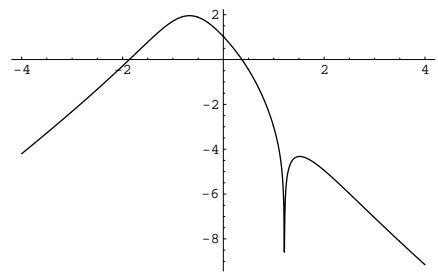}
\caption{Logarithm of the periodic zero mode density for the caloron of
Fig.~\ref{fig_first} as a function of $x_3$ (fixing $x_0=1/2$).}
\label{fig_axis}
\end{figure}

Figure \ref{fig_first} shows the periodic zero mode of an intermediate
caloron with equal mass constituents together with the action
density in the $x_3x_0$-plane.
One can clearly see the two monopole constituents,
of which the zero mode sees only the left one.
Moreover, both quantities exhibit a slight time-dependence.
They are maximal at $x_0=0$ (the center of the instanton lump when
the constituents come even closer) and minimal at $x_0=1/2$.

In Figure \ref{fig_axis} we take a closer look at the periodic zero mode density
at $x_0=1/2$. Beside the asymptotic behavior being an exponential decay, 
a  strong dip in the logarithm shows up around $\bar{x}_3=1.21$.
By zooming into this region, we found
numerically that this dip is arbitrarily deep thus representing a zero (which is
hard to visualize in a plot like Fig.~\ref{fig_first}).
No other zero is present in the profile.

In order to get a handle on this zero analytically, we will now analyze the case
$z=0$ in more detail. 
Due to the property (\ref{eqn_fprop})
the formula for the zero mode simplifies (see also \cite{garciaperez:99c}
using algebraic gauge), 
in particular the different spin components are related as
\begin{eqnarray}
\po(x)_{a2}=-\epsilon_{ab}\,\po^*(x)_{b1}\,.\label{eqn_constraint}
\end{eqnarray}
Hence for the zero it is enough to analyze $\Psi_0(x)_{a1}$.

Furthermore we can neglect $\phi$ and use
that on the axis $x_1=x_2=0$ the derivatives $\partial_1$ and $\partial_2$
vanish
since all $\f$'s depend on $x_{1,2}$ through $x_1^2+x_2^2$ only. This actually
makes the $a=1$ component of $\Psi_0(x)_{a1}$ vanish.

For simplicity we will also specialize to $\omega=1/4$ and $\rho=1$ for the moment.
With the help of {\em mathematica} we found that
$\Psi_0(1/2,0,0,x_3)_{21}$ can (up to non-vanishing factors)
be 
given by the real ratio $N/D_1^2/D_2$, where
\begin{eqnarray}
\v&&N=e^{3\pi x_3}2(2x_3+\pi)\nonumber\\
\v&&+e^{2\pi x_3+\pi^2/2}(-12\pi
x_3^2-4x_3+2\pi+3\pi^2)\nonumber\\
\v&&+e^{\pi x_3}\left[2(2x_3-\pi)+e^{\pi^2}2\pi(2\pi x_3-2-\pi^2)\right]\nonumber\\
\v&&+e^{\pi^2/2}\left[4\pi x_3^2-4x_3(1+\pi^2)+\pi^2(2+\pi)\right],\label{eqn_N}\\
\v&&D_{1,2}=e^{2\pi x_3}(2x_3+\pi)\mp e^{\pi x_3}e^{\pi^2/2}2\pi+2x_3-\pi\,.
\end{eqnarray}
Interestingly, $N$ vanishes at the monopole locations $y_{1,2}=\pm\, \pi/2$,
but these zeroes are compensated by corresponding ones in $D_{1,2}$.
However, $N$ has another zero around $\bar{x}_3=2.06$ where $D_{1,2}$ are positive.
It can easily be shown not to be a numerical artifact since $N$ changes its sign there. 

The zero of the caloron zero mode is located near the `invisible'
constituent monopole (here at $y_2>0$).
This means in particular, that it also hops to the other monopole (at $y_1<0$)
when changing the boundary condition $z$ to around 1/2.
Relative to the monopole location the zero is shifted outwards
(i.e.\ $\bar{x}_3>y_2$),
by an amount which in the two examples above is roughly 0.5.
We have investigated the analogue of Eq.~(\ref{eqn_N}) with arbitrary $\rho$
and $\omega$ and found numerically that for $\omega=1/4$ the shift 
varies indeed from 0.59 for small $\rho$ to 0.46 for large $\rho$.
The dependence on the holonomy is such that the shift vanishes for $\omega=0$
and diverges for $\omega\rightarrow 1/2$ (i.e. $\bar{\omega}\rightarrow 0)$.

As already mentioned above this zero is not accidental.
This can be confirmed by the behavior of the components around it.
We find that the $4\times 4$ gradient matrix consisting of the partial derivatives
 of the four real components of $\Psi_z(x)_{a1}$ has a non-vanishing
determinant at the zero $\bar{x}$. 
Thus around the zero one can normalize the zero mode
which then is a mapping from a little three-sphere in space-time $S^3_{\bar{x}}$
to a three-sphere in internal space $S^3\in\mathbb{C}^2$. 
In particular, this mapping has unit winding number, 
hence is topologically equivalent to a hedgehog (in four dimensions).
This strongly points towards a topological origin of the zero which we discuss now 
beyond the caloron solutions considered so far.

\section{Topological origin: beyond solutions}

In this section we analyze the chiral zero mode in arbitrary gauge field backgrounds
with unit instanton number.
It is always subject to the
charge conjugation symmetry (\ref{eqn_constraint}),
if the mode is periodic or antiperiodic:
with a solution $\p(x)_{a\alpha}$ also 
$\p'(x)_{a\alpha}=\epsilon_{\alpha\beta}\,\epsilon_{ab}\p^*(x)_{b\beta}$
solves (\ref{eqn_eqn}),
where $\bar{\sigma}_\mu^*=\epsilon\,\bar{\sigma}_\mu\epsilon^T$ has been used.
For $z=0$ and $z=1/2$ -- and only for those cases -- 
the function $\p'$ has the same periodicity as $\p$
and thus has to be identical (up to a phase which, however, can be brought to 1).
Therefore, the number of degrees of freedom for the fermion
reduces to two complex ones,
just like for bosons in the fundamental representation
(actually, for the bosonic field of the Laplacian gauge \cite{vink:92}
the charge conjugation
symmetry manifests itself as a two-fold degeneracy \cite{vink:95}).
This second reduction is crucial for arguments presented later,
since otherwise the normalized zero mode would take values on spheres higher than $S^3$ 
and for those the third homotopy group $\pi_3$ is trivial.
What happens for $z \not\in\{0,1/2\}$ is that 
the different spin components do have zeroes of topological origin,
but not necessarily at the same locations
such that their sum need not vanish (we have observed this phenomenon for the caloron).

The zero mode inherits its topology from the gauge field.
The correct mathematical description of the latter is only a local one (see below),
however, for space-times $\mathbb{R}^4$ and $S^1\times \mathbb{R}^3$
it is convenient to work with one global gauge field possessing a gauge
singularity.
The latter is a discontinuity in the gauge field invisible in gauge invariant
quantities.
To be precise, the gauge field around that point can be made smooth upon acting
with a `large' gauge transformation. This is a mapping from a three-sphere into
the gauge group, the winding number of which equals the topological charge.

In the same way the $z=0$ zero mode density is smooth around the gauge singularity,
while the normalized mode
\begin{eqnarray}
\ppo(x)_{a1}=\frac{\po(x)_{a1}}{|\po(x)|}\,,\:
|\po(x)|=\sqrt{\sum_a |\po(x)_{a1}|^2}\,,
\end{eqnarray}
carries a winding number $\pi_3(S^3)$ there.
Sweeping over the whole space-time manifold, the zero mode has to unwind
somewhere, which in a smooth way can only happen at a zero.
Therefore, the only way not to have a zero is to carry the winding up to
infinity
(where the zero mode unwinds at a `zero' that is already demanded by
normalizability).
For the spherically symmetric zero mode of the instanton this is indeed the case.
However, this possibility is excluded for gauge fields over $S^1\times \mathbb{R}^3$
as long as the holonomy is non-trivial!

In order to prove this fact we introduce an auxiliary matrix-valued field
\begin{eqnarray}
g(x)=\left(\begin{array}{cc}
\ppo(x)_{a1}&\ppo(x)_{a2}
\end{array}\right)\,,
\label{eqn_g}
\end{eqnarray}
(resembling the way the zero mode is described in the Sp(1) language) which
is well-defined outside zeroes of $\Psi$ and is an element of $SU(2)$
due to (\ref{eqn_constraint}).
The winding number of the zero mode is most easily expressed as
\begin{eqnarray}
{\rm deg}[\hat{\Psi};\mathcal{S}]=\frac{1}{24\pi^2}\int_\mathcal{S}{\rm d}^3\sigma\,
\hat{n}_\mu\epsilon_{\mu\nu\rho\sigma}{\rm Tr} \bar{A}_\nu\bar{A}_\rho\bar{A}_\sigma\,,
\end{eqnarray}
where $\hat{n}_\mu$ is the normal vector at the three-dimensional surface
$\mathcal{S}$ and $\bar{A}_\mu=g\,\partial_\mu g^\dagger$.

Without a zero the winding persists
over arbitrarily large spatial volumes
\begin{eqnarray}
k\!\!&=\!\!&{\rm deg}[\hat{\Psi};S^3_{\bar{x}}]={\rm deg}[\hat{\Psi};\nonumber\\
&\!\!&B^3_{x_0=0,|\vec{x}|\leq R}\cup
B^3_{x_0=1,|\vec{x}|\leq R}\cup
S^1_{x_0}\times S^2_{|\vec{x}|=R}]\,.
\end{eqnarray}
The first two
contributions coming from three-balls cancel each other since $\Psi$ and hence $g$ and
$\bar{A}_\mu$ are periodic and $\hat{n}_\mu=(\pm 1,0,0,0)$ has the opposite
sign.

The last term is proportional to $\bar{A}_0$, which for large enough $R$ will
vanish:
acting with $\sigma_\nu(\partial_\nu+A_\nu)$ on Eq. (\ref{eqn_eqn}) gives that
the zero mode asymptotically is subject to the Klein-Gordon equation
$\left((\partial_0+2\pi i\omega\tau_3)^2+\vec{\partial}^2\right)\po=0$.
Here we have neglected the decaying fields $A_i$ and $F_{\mu\nu}$ and used that
$A_0$ approaches the logarithm of $\P$ (in periodic gauge).
The decomposition of $\po$ into a Fourier series $\exp(2 \pi i p x_0)$ results
in constant potentials $4\pi^2(p\pm \omega)^2$ for the spatial Laplacian.
Since the height of the potential gives the spatial decay rates and $\omega\leq
1/2$,
one needs to look at the $p=0$ component for the leading order.
Therefore $\po$ and $g$ are
time-independent and $\bar{A}_0=0$ which leads to a vanishing topological charge
$k$.

The conclusion is that {\em the periodic and antiperiodic zero mode
in backgrounds with non-vanishing topological charge
must have a zero at finite distance}. 
The only exception is $\P=-\Eins_2$ (for $z=1/2$ it is $\P=\Eins_2$) where our
considerations are not valid
(since for $\omega=1/2$ the $p=\mp 1$ components give the same height
of the potential as $p=0$)
and indeed, the periodic caloron zero mode has its zero at spatial infinity
for $\omega=1/2$ (see above).
Notice that the only property we used from the background gauge field $A_\mu$
is its asymptotic behavior which is required by finite action \cite{gross:81}
and non-trivial holonomy.

These topological arguments do not determine the number of zeroes.
Typically, classical solutions such as calorons possess the minimal number of zeroes.
In this respect the zero mode behaves similar to the Polyakov loop.
Topology requires the latter to go through $\pm\Eins_2$, non-trivial holonomy 
prevents this from happening at infinity (which is the case for the instanton) 
and for the caloron there is just one location for each of the two values
\cite{garciaperez:99a}.

\section{Generalization to higher charge and other space-times}

For configurations of {\em higher topological charge} $|k|>1$ our arguments
generically predict $|k|$ zeroes in the zero mode.
The reason is simply that the zero mode has to unwind a higher winding number
$k$, but the change in the latter at a generic zero is only $\pm 1$.

\begin{figure}[h]
\includegraphics[height=2.5cm]
{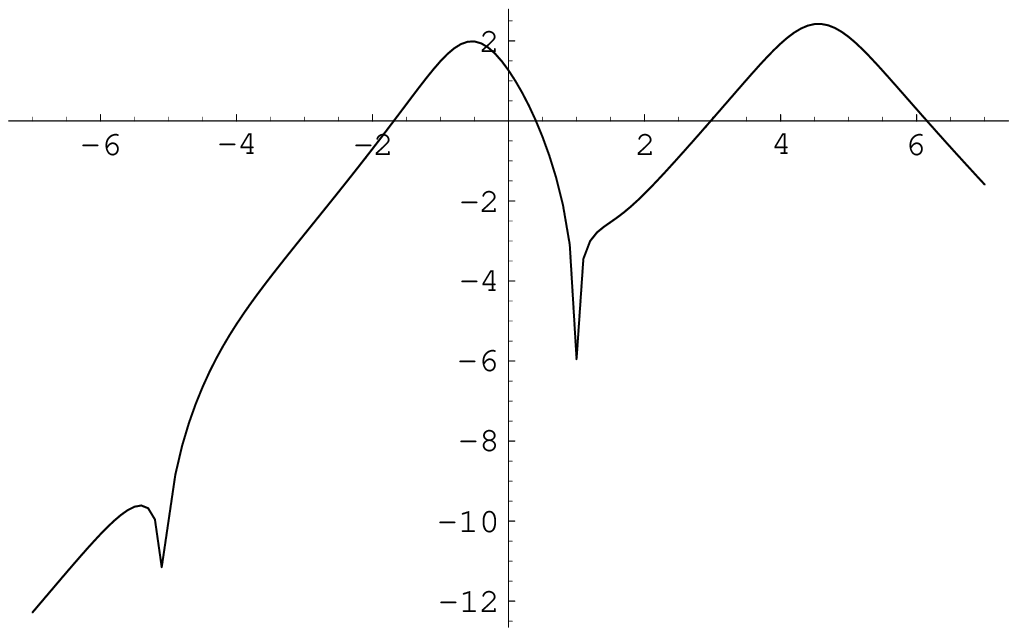}
\includegraphics[height=2.7cm]
{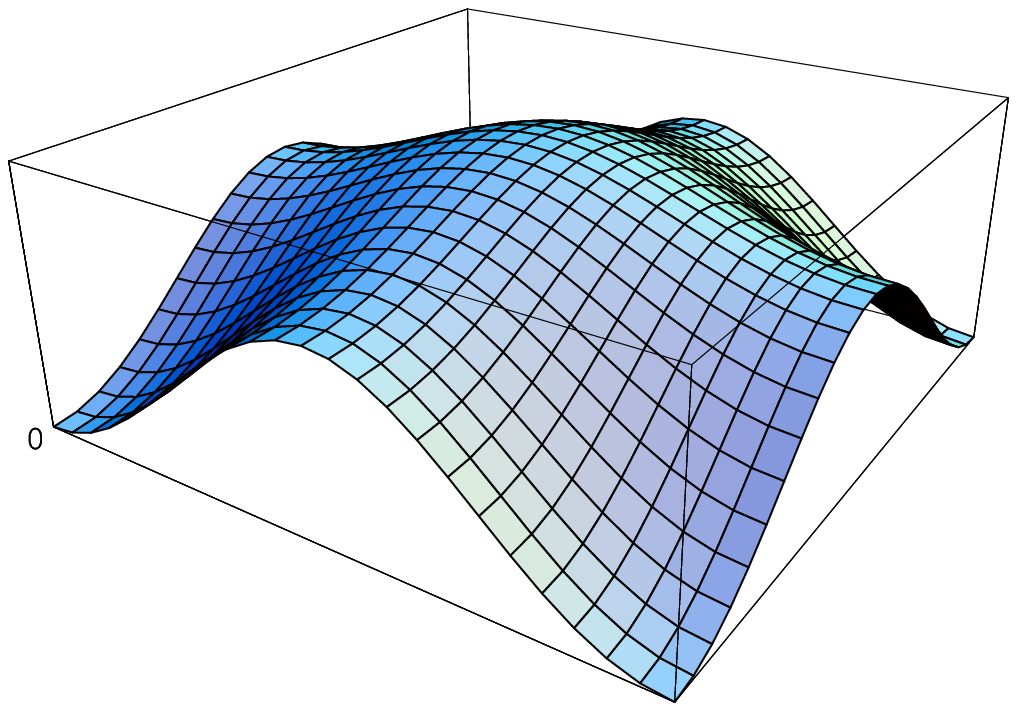}
\caption{On the left is shown the logarithm of a zero mode of an axially
symmetric charge 2 caloron as written down in \cite{bruckmann:02b},
at $x_0=1/2$
(for the action density see the left part of Fig.\ 4 in that reference).
By construction the constituent
monopoles are located at $\{-0.56,\,4.56\}$ and at $\{-4.56,\,0.56\}$, respectively.
Indeed the maxima of the zero mode come at $\{-0.54,\,4.56\}$, while two zeroes occur
at $\{-5.07,\,0.99\}$. On the right is shown the density of a zero mode for an
Abelian instanton of charge two over two coordinates on the four-torus.
The point-like zero visible in the figure extends to a whole two-torus in the
remaining directions.}
\label{fig_charge_two}
\end{figure}

In Figure \ref{fig_charge_two} we show a zero mode of an axially symmetric caloron of
charge two. It illustrates once more that while the maxima of
the mode are related to two monopoles, there are two zeroes in the vicinity of
the monopoles of the other type.

Our findings also hold over {\em compact space-time manifolds},
e.g.\ the four-torus.
In these cases we use that $\Psi$ is defined locally over patches $U_I$ and is
subject to the same transition functions
\begin{eqnarray}
\Psi_J=t_{IJ}^{-1}\,\Psi_I\,,\quad\mbox{no sum,}\label{eqn_trans}
\end{eqnarray}
as the background gauge field, $A_J=t_{IJ}^{-1}(A_I+{\rm d})t_{IJ}$.
Furthermore, we assume that $\Psi$ has no zero such that $\hat{\Psi}$ and $g$
in (\ref{eqn_g}) are well defined and fulfill (\ref{eqn_trans}) as well.
Let us endow the principal fiber bundle with a set of local sections
translating as $s_J=s_I\,t_{IJ}$ \cite{nakahara:90}.
Then the fermion-induced group-valued field $g_I$ can be used to define
a new set of sections $s'_I\equiv s_I\,g_I$ which obviously does not change at
the overlap of patches, $s'_J=s'_I$, and thus is equivalent to a global section.
This renders the bundle trivial, and the topological charge vanishes.

For an alternative proof one can use $\bar{A}$ which is a connection in the
same bundle as $A$ (has the same transition functions) but being a pure gauge
it cannot generate field strength nor topological charge.

We conclude that fermionic zero modes in non-trivial backgrounds always have to
possess a zero.
In simple words, the topology of the background gauge field twists the zero mode
such that it can follow only by virtue of a zero
(in the same spirit as over $S^1\times\mathbb{R}^3$).
Notice that again we needed no further assumptions about the gauge field.
The zero mode equation (\ref{eqn_eqn}) was used only insofar as it reduces the
number of degrees of freedom.
Thus our statements also hold for bosonic fields in the fundamental
representation, by writing $g=(\hat{\phi}_a\:\:-(\epsilon\hat{\phi}^*)_a$).
Additional phases $-1$ (`antiperiodic boundary conditions') are allowed in both
cases.

For an example on the four-torus we take as a background an
Abelian charge two instanton with constant field strength \cite{thooft:81c}.
Its zero mode \cite{vanbaal:96} is shown in Figure \ref{fig_charge_two} 
over two directions revealing a zero.
On such a two-torus holomorphic techniques have been used to show that the
number of zeroes is proportional to the instanton number \cite{aguado:01}.
Due to the fact that the full zero mode is a product of profiles over two-tori,
the zero actually extends over a whole two-dimensional torus.
This is non-generic because the configuration
represents a corner of the moduli space.
From the explicit formulae \cite{vanbaal:96} it is evident that the zero
moves upon changing the boundary conditions.

As for higher gauge groups $SU(N\geq3)$,
the relation between $\p(x)_{a1}$ and $\p(x)_{a2}$
resp.\ the topological arguments will be more involved and in the caloron there
are more constituents to be detected. Results about these more complicated cases will
be reported elsewhere.

\section{Discussion}

We have examined the existence of zeroes in zero modes of non-trivial backgrounds
for both finite temperature and compact space-times by 
topological arguments.
By means of the caloron we have also demonstrated that the zero location is related to
the constituent monopoles, although the zero is not line-like but has
codimension four by construction.

The constituent picture on the
four-torus is not fully understood. Still moduli space arguments, the existence
of fractionally charged instantons (with twist \cite{thooft:79}) and numerical
investigations \cite{gonzalez-arroyo:95,bruckmann:04b} suggest $2k$
constituents, however not strongly localized.
The zero mode zeroes might help here to reveal the locations of them.
Compared to the caloron setting the finite volume actually has the advantage
that the periodic copies of the constituents should cure the shift discussed
above (`push the zero back inwards'). 

The zero mode can also be used to fix the gauge.
Using $g$ as the
gauge transformation which does the gauge fixing results in
defects -- points where the gauge fixing is not well defined --
exactly at the zeroes.
For the same reason defects necessarily occur in the Laplacian gauge as well.

Defects are substantial to Abelian projections \cite{thooft:81a}.
An Abelian gauge
can easily be defined by gauge fixing with $g\exp(i\alpha\tau_3)$.
This amounts to diagonalizing the operator $g\tau_3 g^\dagger$ transforming in
the adjoint representation.
By construction this composite operator has point-like defects
as opposed to a generic adjoint operator which has line-like defects,
thin monopoles.
Furthermore, as shown above, this operator possesses defects even for Abelian
(so-called reducible) backgrounds, while not all such operators need to do
so, for instance $\tau_3$ is free of defects and also subject to the correct topology
(Abelian transition functions).

We propose to use our findings in lattice configurations, namely to
uncover the topological/infrared content of them.
Due to the filter property of the zero mode one should be able to detect the
zero by interpolating the components (and using their winding around the zero).
If necessary, smeared backgrounds can be used for this purpose.
Preliminary results on calorons obtained by cooling (cf. \cite{ilgenfritz:02a})
are promising.

It would actually be very interesting to use the bosonic field of the Laplacian
gauge as a filter, too. That is to introduce nontrivial phase boundary
conditions for this field and to look how its maxima (and zeroes which have been
connected to instantons \cite{bruckmann:01a,deforcrand:01a})
change under them and how the latter are
related to the corresponding maxima and zeroes in the fermion zero mode.
Since the bosonic operator $-(\partial_\mu+A_\mu)^2$
 has less components and no chirality issues, it should
be computationally cheaper.

\section{Acknowledgments}

The author thanks Christof Gattringer, Daniel Nogradi and Pierre van Baal
for helpful discussions and reading the manuscript
and Ernst-Michael Ilgenfritz and Dirk Peschka for providing lattice data.
The author has been supported by DFG and in an early stage by FOM.

%\bibliographystyle{../revtex4/apsrev}
%\bibliography{../../gauge}
\bibliography{}   

\end{document}